# SIM-STEM Lab: Incorporating Compressed Sensing Theory for Fast STEM Simulation

Alex Robinson[1], Daniel Nicholls[1], Jack Wells[2], Amirafshar Moshtaghpour[1,3], Angus Kirkland[3,4], Nigel D. Browning[1,5,6]

[1] Department of Mechanical, Materials and Aerospace Engineering, University of Liverpool, Liverpool, L69 3GH, United Kingdom

[2] Distributed Algorithms Centre for Doctoral Training, University of Liverpool, Liverpool, L69 3GH, United Kingdom

[3] Rosalind Franklin Institute, Harwell Science and Innovation Campus, Didcot, OX11 0QS, United Kingdom

[4] Department of Materials, University of Oxford, Oxford, OX2 6NN, United Kingdom

[5] Physical and Computational Science Directorate, Pacific Northwest National Laboratory, Richland, WA 99352, USA

[6] Sivananthan Laboratories, 590 Territorial Drive, Bolingbrook, IL 60440, USA

## Abstract

Recently it has been shown that precise dose control and an increase in the overall acquisition speed of atomic resolution scanning transmission electron microscope (STEM) images can be achieved by acquiring only a small fraction of the pixels in the image experimentally and then reconstructing the full image using an inpainting algorithm. In this paper, we apply the same inpainting approach (a form of compressed sensing) to simulated, sub-sampled atomic resolution STEM images. We find that it is possible to significantly sub-sample the area that is simulated, the number of **g**-vectors contributing the image, and the number of frozen phonon configurations contributing to the final image while still producing an acceptable fit to a fully sampled simulation. Here we discuss the parameters that we use and how the resulting simulations can be quantifiably compared to the full simulations. As with any Compressed Sensing methodology, care must be taken to ensure that isolated events are not excluded from the process, but the observed increase in simulation speed provides significant opportunities for real time simulations, image classification and analytics to be performed as a supplement to experiments on a microscope to be developed in the future.

## Acknowledgements

*This work was performed in the Albert Crewe Centre (ACC) for Electron Microscopy, a shared research facility (SRF) fully supported by the University of Liverpool. This work was also funded by the EPSRC Centre for Doctoral Training in Distributed Algorithms (EP/S023445/1) and Sivananthan Labs. The authors would also like to recognise the efforts of Ivan Lobato et al. for their development of MULTEM, without which this research would have not been*

## 1. Introduction

The simulation of atomic resolution scanning transmission electron microscopy (STEM) images [1- 4] has advanced significantly in recent years, to the point that the methodology is now capable of identifying single atom changes in structure and composition from a direct comparison to experimental images [5- 8]. Simulations are now used to help understand the structure-property relationships in a wide range of beam tolerant materials, interfaces, grain boundaries and defects [9- 18]. These simulations, however, typically involve computationally expensive formalisations where the core underlying approach is to include as many possible contributions/configurations as possible to ensure the best match to experimental images [19-39, 53, 54]. In addition to the significant computational power required and run time necessary to make sure the simulations have converged to best match the experiment; this approach runs into severe limitations when the comparison experimental image quality is poor [40- 44] . This means that if a material, interface, or defect is susceptible to electron beam damage, it is hard to develop precise simulations of these structures.

For these damage limited acquisitions, there are currently two approaches to improving the inherent low signal-to-noise ratios that are being applied that make use of artificial intelligence: one is by using compressive sensing/inpainting [45- 47] and the other is by classifying and learning features from a large number of low-quality images [48- 50]. Could something similar be applied to simulations?  Rather than performing one perfect simulation, could we run a set of smaller less precise (sub-sampled) simulations and use machine learning/artificial intelligence to classify and quantify the structure for comparison to experimental results?  Even further, if we use machine learning to classify images, is there really a difference between the use of a simulated image and an experimental image to solve the structure at hand?

In this paper we will focus only on one use of artificial intelligence for image/simulation reconstruction, sub-sampling and inpainting, i.e., compressive sensing (CS), which has been shown to reduce the time and electron dose rate needed to form STEM images [40- 42]. Compressive sensing STEM involves forming images that are directly subsampled at the time of acquisition (Figure 1). An inpainting algorithm is then used to fill in gaps in the sub-sampled data, with missing information inferred from the subsampled data through, e.g., a combination of a dictionary learning or a sparsity pursuit algorithm (Figure 2). To estimate a full image from simulated subsampled image, we use an unsupervised dictionary learning method. Compared to its supervised counterparts, which require a suitable pre-defined dictionary. Dictionary learning algorithms produce a dictionary of basic signal patterns, which is learned from the data, through a sparse linear combination with a set of corresponding weights. This dictionary is then used in conjunction with a sparse pursuit algorithm to inpaint the pixels of each overlapping patch which when combined form a full image [51]. In this paper, the Beta-Process Factor Analysis via Expectation Maximisation (BPFA-EM) method will be used to inpaint the subsampled data, and the reader is referred to [52] for more detail. We do not claim that BPFA is the state-of-the-art blind inpainting algorithm, however it is one of the top performing models outside of deep-learning methods. BPFA does not require prior knowledge of measurement noise, unlike its counterpart, i.e., K-SVD [53]. This is beneficial for real CS-STEM where noise level estimation is not always known, but for these simulations, noise is not considered prior to inpainting.

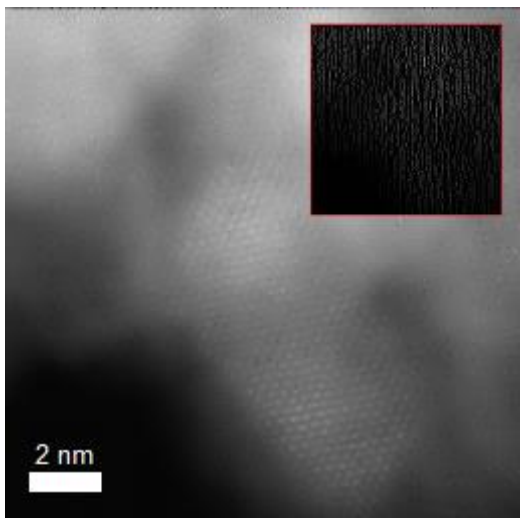

**Figure 1**. An example of a BPFA-EM reconstructed CS-STEM image of heat-treated Ceria. The overlayed image is subsampled by 8x acquired data (i.e., only 12.5% of scan positions acquired) using a 'line-hop' sampling scheme. This theoretically reduces the total electron counts (and equivalently the dose for constant dwell time and beam current) by 8x. Given the success of the method in real CS-STEM, it is hypothesised that the same approach can be applied to STEM simulation.

To determine whether it is possible to use the same inpainting approaches to the generation of image simulations, we first must examine what is involved in simulating atomic resolution images. In this regard, the multislice approach [54, 55] is a common method for simulating the projected electron wavefunction, where the three-dimensional atomic potential is divided into a series of two-dimensional slices. The weak phase object approximation (WPOA) is assumed where only the phase of the incident electron wave is slightly modified [34], but not the amplitude. To account for electron-phonon interactions, the frozen phonon model is often used [33- 35]. The intensity average of a series of multislice calculations is taken for different possible configurations of atomic coordinates (Figure 5), and the number of frozen phonon configurations is linear with the run-time. In STEM simulations, this series of calculations must be solved at each probe position, which ultimately defines the resolution of the resulting simulation. Therefore, the run-time of STEM simulation scales not only with resolution, but the theoretical accuracy of the desired calculation, with a 128x128 pixel simulation taking on the order of $1 \times 10^4$ seconds with only 10 frozen phonon configurations, and a sample thickness of

1nm. Typically, the user must compromise between the run-time, accuracy, and size of the simulated image or must invest in high-end machines which can speed up the process through GPU parallelization [20-25], such as that used by the MULTEM software referenced and used in this paper. It is in overcoming this compromise between run-time and accuracy, that inpainting can provide the most benefit.

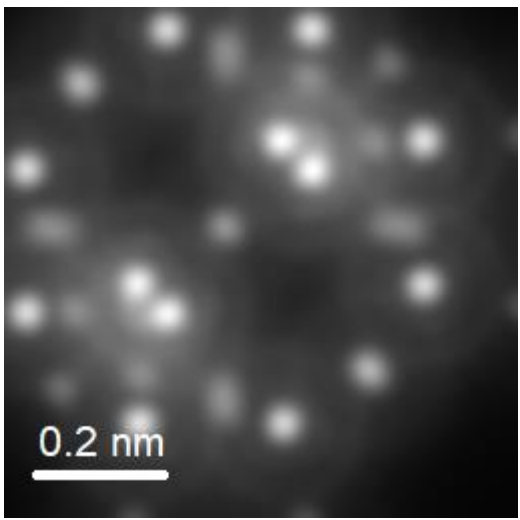

**Figure 2**. An example of a simulated image of ZK-5 zeolite. The simulated image is a HAADF STEM simulation with an accelerating voltage of 300kV on a detector with inner and outer angles of 60 mrad and 100 mrad, respectively. The simulation used 32 frozen phonon configurations, and a maximum reciprocal space vector of $10.92 \text{ Å}^{-1}$. The time-to-solution was 34,154.56 seconds.

In this paper, we will discuss how combining inpainting with STEM simulations, it is possible to form a new workflow for structural determination which can significantly reduce the run-time required for functionally identical results. This development allows real-time simulations to be performed, which is the first step in the rapid interpretation, classification, and analysis of images and potentially the future development of artificial intelligent (AI) STEM, AISTEM.

## 2. Methods

In the following section, three possible methods to reduce the run-time of STEM simulations will be presented and explained. The first is based on probe subsampling, the second is related to the number of frozen phonon configurations (FPC) used, and the third is based on the optimisation of the maximum reciprocal space vector that contributes to the simulated image. In all cases, the simulations were performed using MULTEM through MATLAB on a remote server equipped with an Intel Xeon Gold 6128 CPU @ 3.40 GHz, and one NVIDIA Telsa V100 GPUs running CUDA 11.2. All simulations are performed with the same microscope and detector parameters. The simulated images are HAADF STEM simulations with an accelerating voltage of 300kV on a detector with inner and outer angles of 60 mrad and 100 mrad respectively, and details on the number of frozen phonon configurations used can be found in the sections below.

## 2.1. Subsampled Area Simulations

It is possible to directly subsample a simulation using the MULTEM scripts in MATLAB by a method of discrete patch simulation. A simulated image of size $[M \times N]$ pixels is constructed from a series of simulated patches of size $[a \times b]$ pixels where $a, b \in [2, min\{M, N\}]$, as is demonstrated in figure 3. The minimum patch size in the current implementation of this method is [2 x 2] pixels, hence the lower bound of a, b. The number of patches required to be simulated is defined by the desired sampling percentage. There are several possible schemes for subsampling those patches, such as 'random sampling', 'line hop sampling', 'adaptive sampling', 'Poisson disk sampling', 'spiral sampling', 'radial sampling', and so on. Given that this method is not limited by beam damage or other acquisition effects, a sampling pattern (or patch selection set) could be designed in such a way that reconstruction quality is maximised without any computational cost. As the results will all follow approximately the same trajectory of quality vs sampling percentage, here we will just focus on a single random scattering sub-sampling pattern.

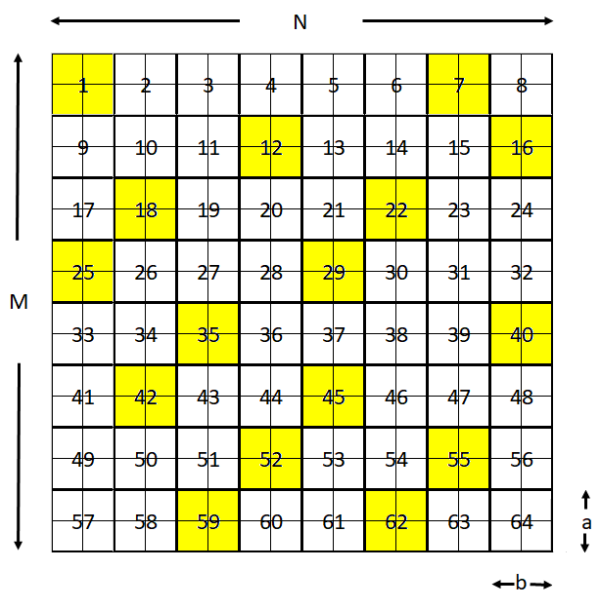

**Figure 3**. A pre-determined image size is defined as [M x N] pixels. This space is then divided into simulation patches of size [a x b] pixels (bounded by thick lines). These patches are then indexed, in this case the set [1,64]. Depending on the desired sampling scheme, a subset of these patches (in yellow) is simulated independently, and then restored into their respective position on the total output image. This then produces a subsampled simulation, as seen in figure 4.

MULTEM asks the user to define the scan area in terms of Angstroms, hence when simulating a patch, the area of that patch is called, simulated, and then saved into a matrix of simulated data. The resulting image is the collection of simulated patches restored in their respective position, with patches that have not been simulated set at a value of zero (Figure 4). Once the subsampled simulation has been generated, it is then reconstructed using the BPFA-EM algorithm (figure 4).

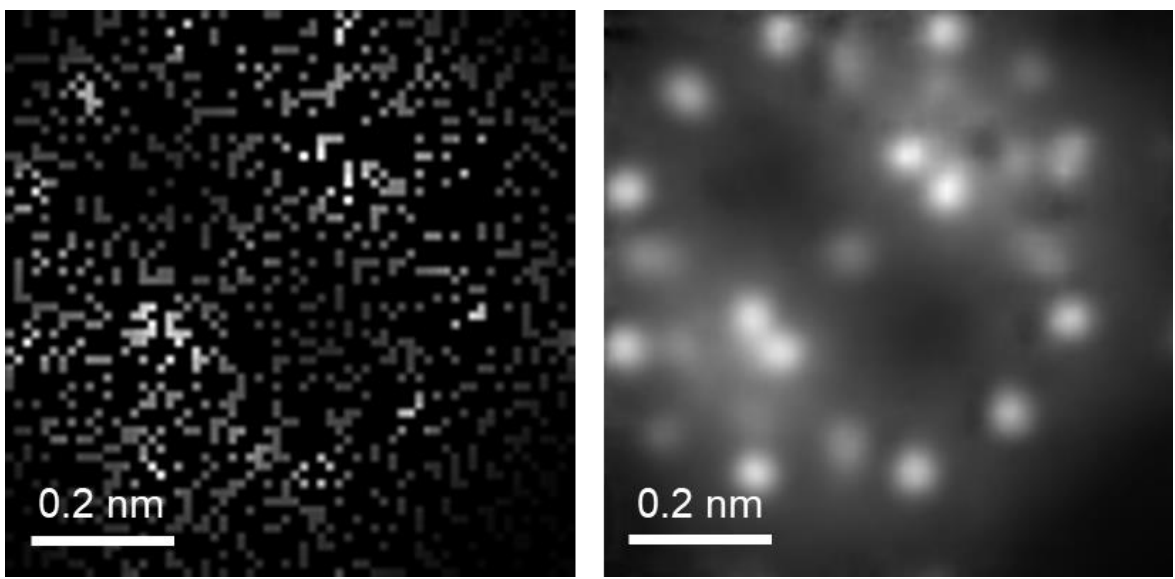

**Figure 4.** An example of a directly subsampled simulation of ZK-5 zeolite (left). The simulation used 5 frozen phonon configurations, and a maximum reciprocal space vector of $5.46\,\text{Å}^{-1}$. The time-to-solution was 479.74 seconds, with approximately an extra 16 seconds to reconstruct the subsampled data using BPFA-EM (right).

## 2.2. Optimising the Number of Frozen Phonon Configurations

The second method to decrease the run-time of STEM simulations (which can be used in conjunction with the subsampling method) is to reduce the number of frozen phonon configurations. For thicker samples, the variance in the output image is reduced given there are more slices to calculate and to average out. There exists a number of configurations beyond which the gradient of the image improvement falls below one, i.e., diminishing returns for increasing number of configurations. This number is inversely proportional to the thickness of the sample. Hence, if the number of configurations is sufficiently large given the thickness of the sample, the output image is functionally similar to the output for many more configurations. It is important to note that this manipulation is intended to increase speed and not accuracy and for complex samples containing defects or dopants, a larger number of configurations is generally recommended to account for the larger atomic position uncertainty.

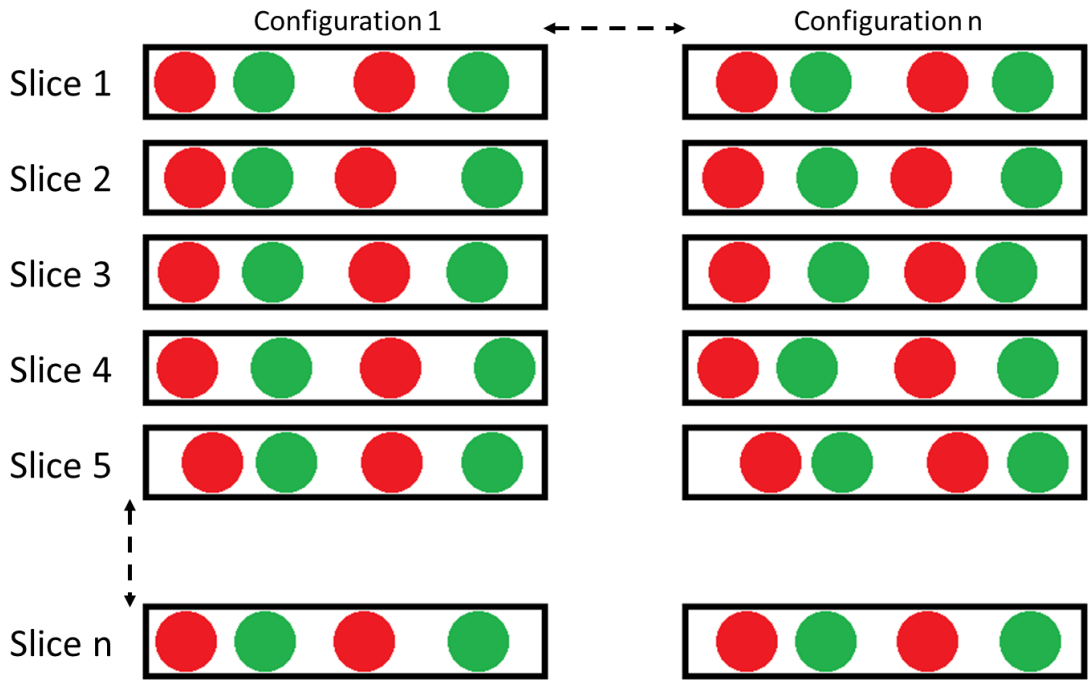

**Figure 5**. A visual representation of how the frozen phonon approximation is used to estimate the thermal-diffuse scattering. For each slice of the multislice, the atomic displacement is randomly assigned for n-configurations. The intensity is calculated for each configuration, and then averaged to determine the intensity of the final image.

## 2.3. Optimising the Maximum Reciprocal Space Vector

Similar to reducing the number of frozen phonon configurations, the maximum reciprocal space vector, or the simulation space, which contributes to the final solution in the simulation can be reduced (Figure 6). Optimising this number reduces the number of calculations required per multislice calculation, and hence reduces the total run-time of the simulation. For a radial detector with an outer angle of θ, the maximum reciprocal space vector (**g**-max) that is incident on the detector will be approximately $\frac{\theta}{\lambda}$ where λ is the relativistic wavelength of the electron. Therefore, we can calculate the maximum simulation box size required to cover the entire detector (figure S3), reducing the number of calculations required, and speeding up the simulation.

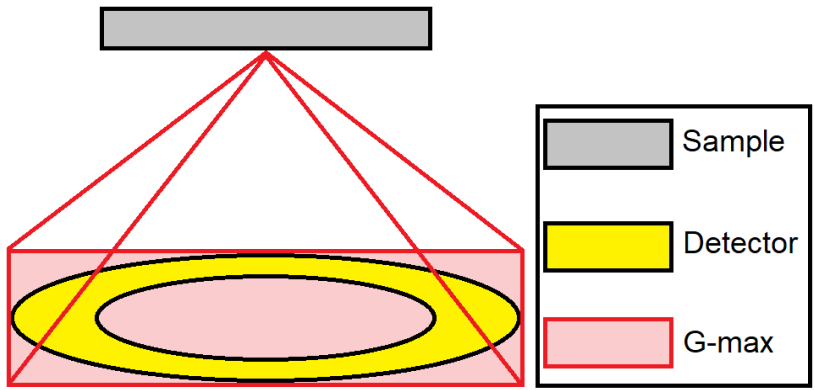

**Figure 6**. A visual representation of how the maximum reciprocal space vector can be interpreted. If **g-max** is less than the corresponding outer angle of the detector, then incoherent signals will be missed for the output. However, if g-max is too large, the incoherent signal outside the detector space is negligible and therefore the run-time is compromised.

## 3. Results

The methods above were tested using the ZK-5 zeolite sample shown in figure 2. The patch size used for all sampling ratios is $[2 \times 2]$, and the image size is 128 x 128 pixels (8Å x 8Å). The metrics used in each of the methods are the structural similarity (SSIM) [56], and peak signal-to-noise ratio (PSNR) [57]. The performance of the methods is summarised in figures 7, 8 and 9. The reference simulation used 32 frozen phonon configurations, and a maximum reciprocal space vector of $10.92\,\text{Å}^{-1}$, and the time-to-solution was 34,154.56 seconds. Each simulation was performed 10 times for each method parameter (sampling ratio, number of configurations, and maximum reciprocal space vector) to get an average image quality metric value as shown in the plot. The error bars correspond to one (plus/minus) standard deviation.

The run-time of the simulation scales linearly with the sampling percentage, however beyond approximately 20% sampling, the increase in quality of image output decreases, causing diminishing returns in image quality with increased sampling percentage. It is therefore proposed that even given this sub-optimised sampling regime and patch size, we can achieve a PSNR value greater than 32dB (on average). Although the patch size used is somewhat inefficient during the simulation, given that the time it takes to transfer data between GPU and CPU memory versus the run-time per patch is the greatest (see figure S5), the smaller patch gives greater sparsity in the acquisition model and therefore greater reconstruction quality, implying greater efficiency to final solution given smaller patch sizes.

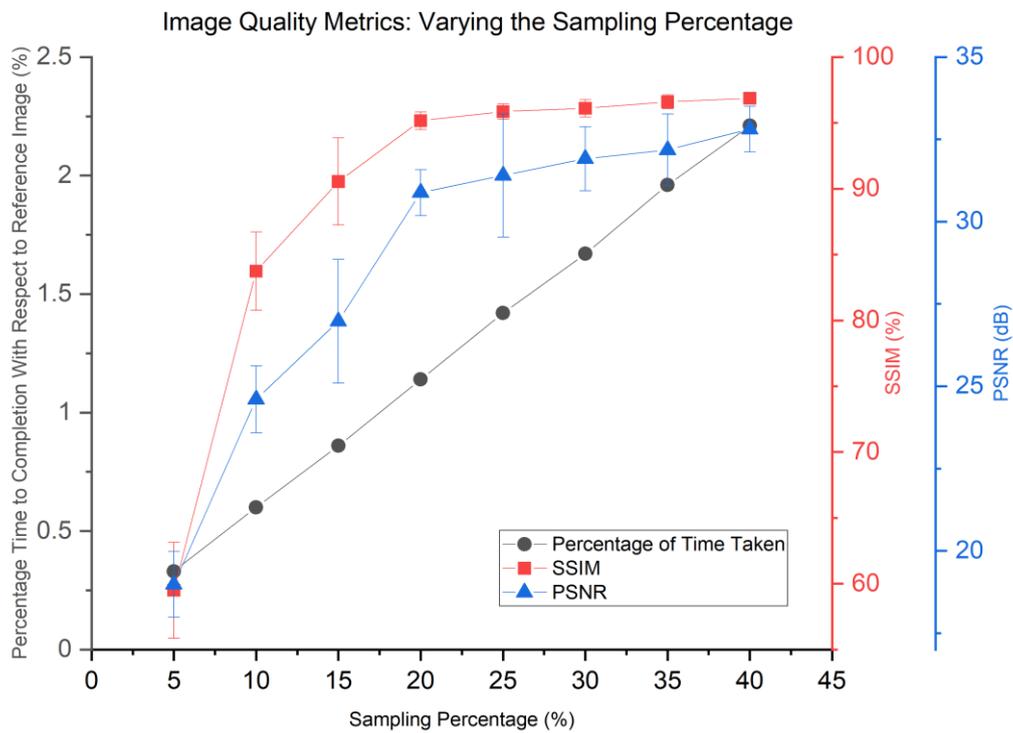

**Figure 7.** The effect of sampling percentage on reconstruction quality. In these simulations, a maximum reciprocal space vector of $5.46\,\text{Å}^{-1}$ was used, and 5 frozen phonon configurations. These parameters were determined sufficient for the specific sample used.

The run-time is linear to the number of frozen phonon configurations (Figure 8), but beyond approximately 5 frozen phonon configurations, the increase in simulation reconstruction quality (with respect to the reference in figure 2) begins to diminish, and as such is determined the optimum for this sample. Of course, for more complex samples containing grain boundaries or defects, one would require more configurations to account for the uncertainty of atomic displacements. Furthermore, for thinner samples it is expected that more frozen phonon configurations are beneficial.

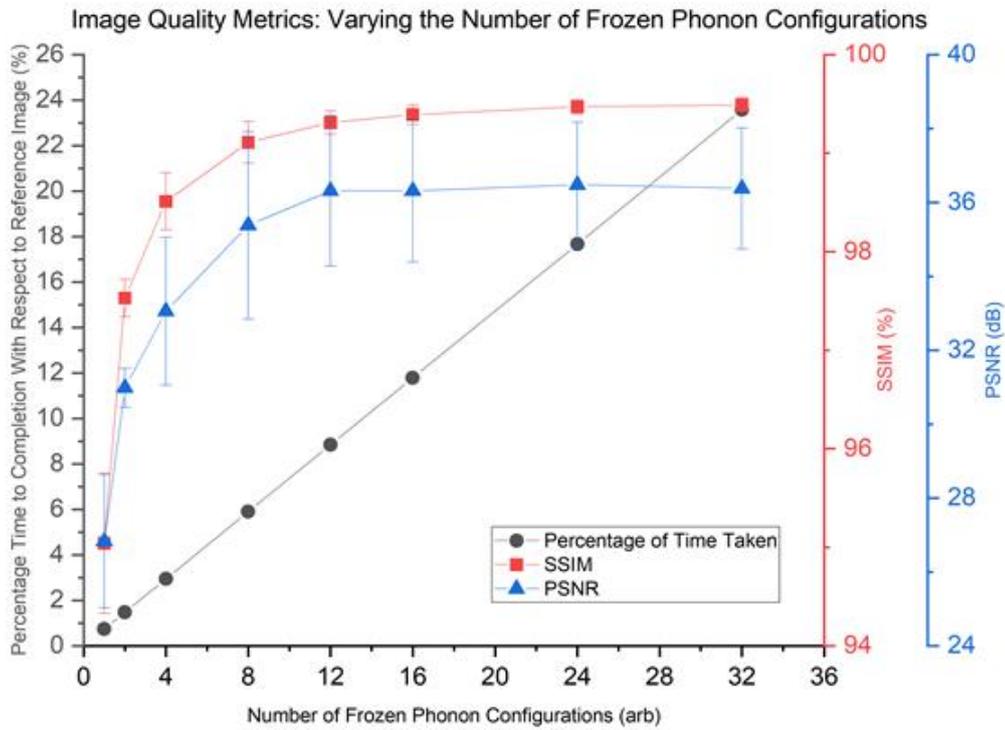
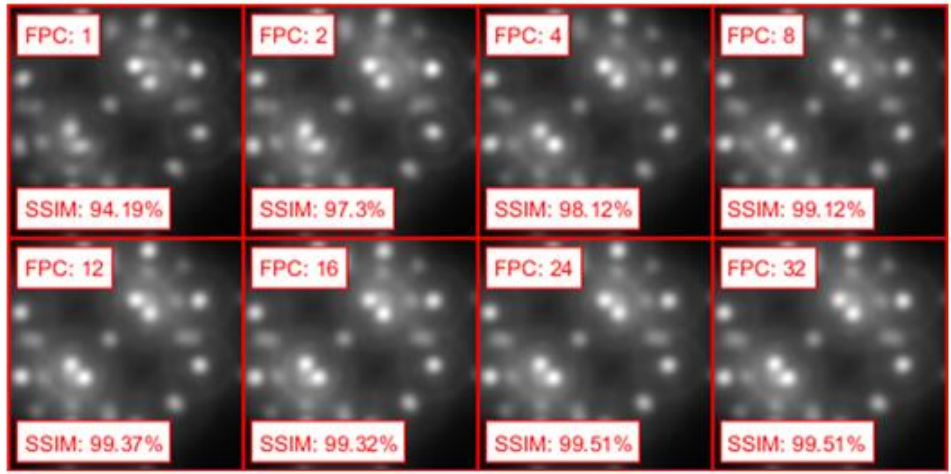

**Figure 8**. (top) The result of increasing the number of frozen phonon configurations upon run-time and image quality metrics. The maximum reciprocal space vector considered for this series is $5.46 Å^{-1}$, and all simulations were fully sampled. (bottom) Examples of simulations performed with varying numbers of FPCs.

As seen in figure 9, there exists a certain maximum reciprocal space vector beyond which the return in image quality increase becomes negligible, and any calculations beyond this not only yield no significant quality increase but take a disproportionally increasing amount of run time. The run-time is generally proportionate with the square of the maximum reciprocal space vector. The HAADF signal is considered a dominantly incoherent signal, and as a result, the electrons on the detector have little phase interference with the electrons that scatter beyond the detector. Therefore, it is expected that to sample electrons beyond the size of the HAADF detector is computationally inefficient, as demonstrated in figure 9.

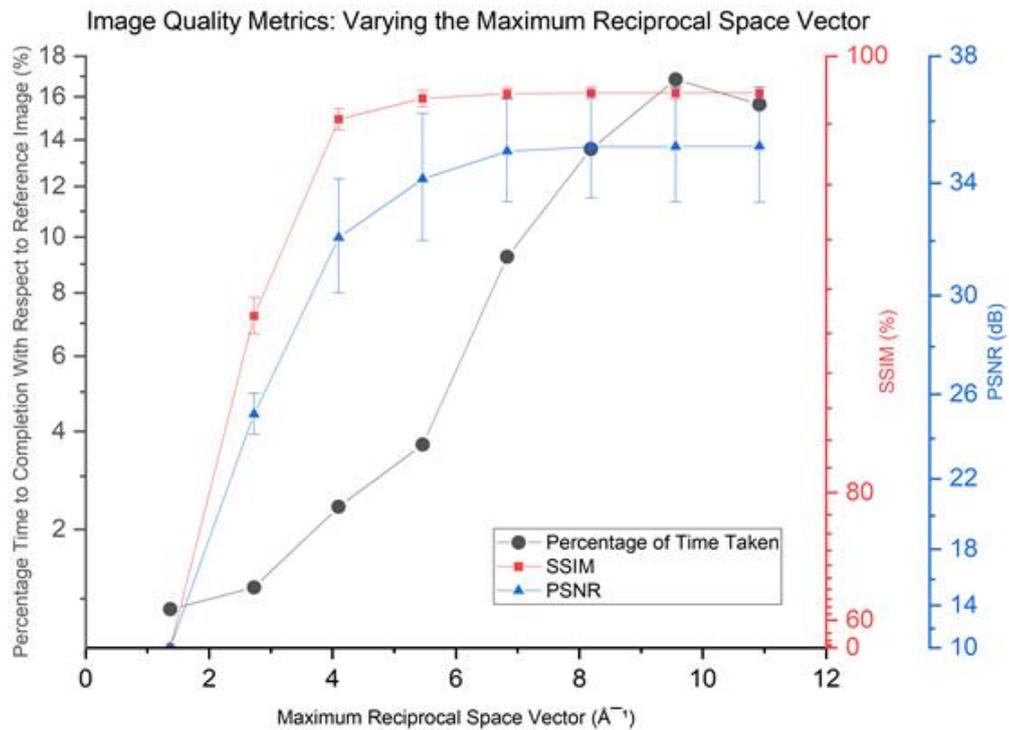

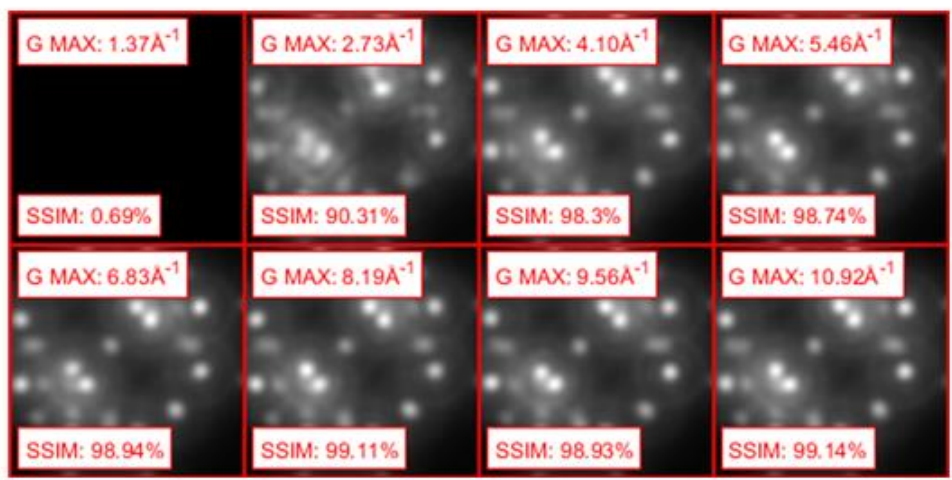

**Figure 9**. (top) Demonstration of how the increase in the maximum reciprocal space vector used has diminishing returns on the output image. Beyond the length of the simulation box being equal to the outer diameter of the detector, the simulations are functionally identical, the size of this box being 5.12x5.12 Å$^{-1}$ (also see figure S2). The simulations were run with 5

frozen phonon configurations, and all were fully sampled. Note the non-linear axes. (bottom) Examples of simulations performed with varying maximum reciprocal space vectors.

In the next experiment, all three methods were used in conjunction to simulate a zeolite structure (ZK-5). The output simulation has a size of 128 x 128 pixels (8Å x 8Å) and was simulated with a sampling percentage of 25% over a random sampling pattern. The number of frozen phonon configurations per slice was 4 and the simulation box was limited to the size of the detector with a maximum reciprocal space vector of $5.12Å^{-1}$. The resulting compressed simulation (figure 10 (c)) has a structural similarity of 95.59% and a peak signal-to-noise ratio value of 29.09dB with respect to the fully sampled simulation. The run-time was approximately 87x faster than the fully sampled simulation, and reconstruction time was on the order of 10 seconds, with most of this time spent forming the dictionary as opposed to actual reconstruction. One could therefore save time by reconstructing using a pre-learned dictionary which is suitably representative of the sub-sampled data.

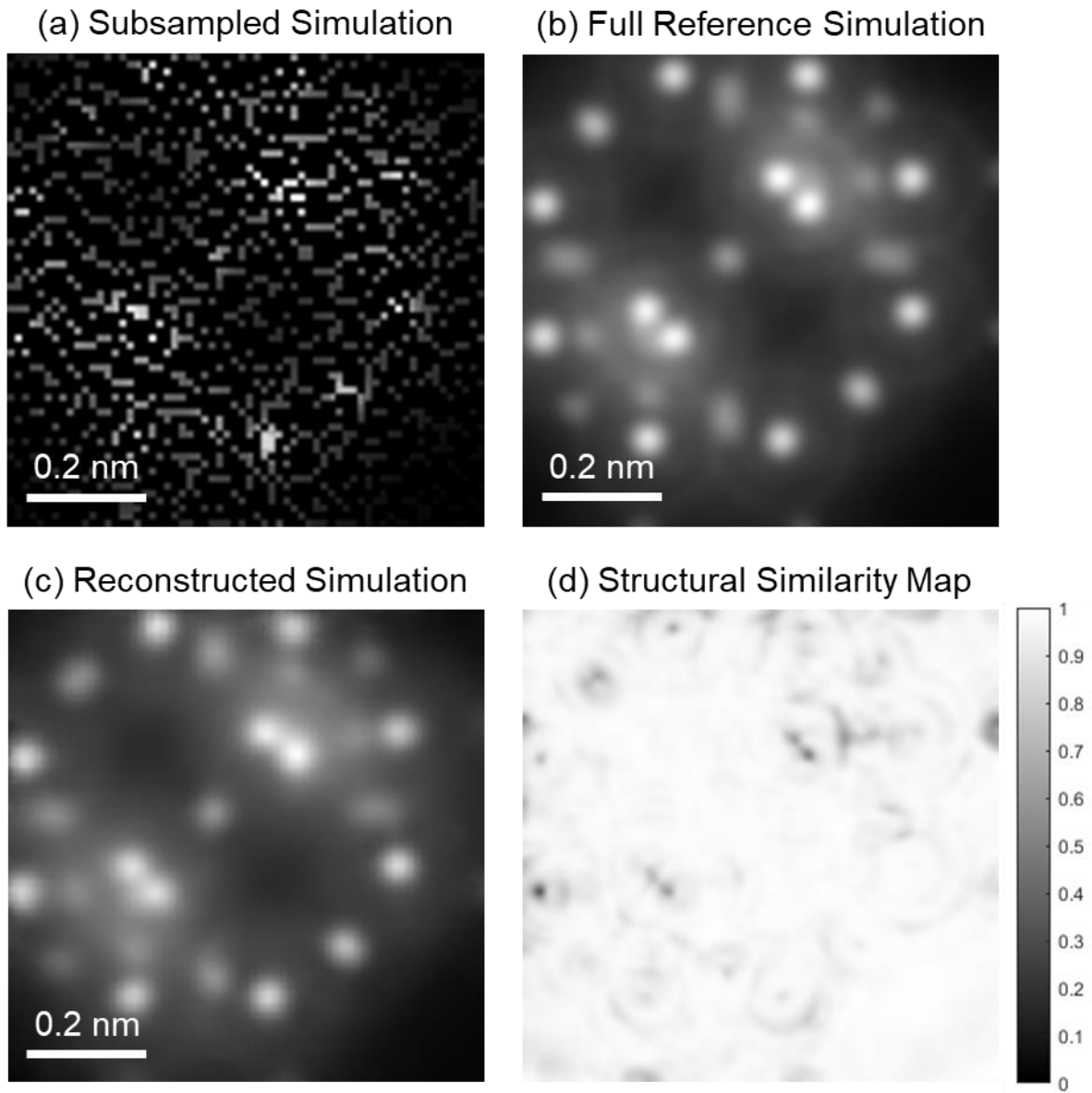

**Figure 10**. (a) Subsampled simulation, (b) the full reference simulation, (c) reconstructed simulation from subsampled data (a) and (d) structural similarity map of the ZK-5 compressed simulation with respect to the reference. In the similarity map, the dark features correspond to larger errors between the reference and reconstruction. In a perfect reconstruction (i.e., SSIM is 1 or 100% similar), the similarity map would be completely white.

## 4. Conclusion

In conclusion, we have shown that it is possible to significantly reduce the run-time of STEM simulation by leveraging the theory of compressed sensing. Furthermore, the results of using these three methods in combination yield results that are functionally similar, often identical to the fully simulated images. It is important to note that these methods are not expected to yield greater accuracy but are to be used in conjunction with experimental results, which are themselves inherently imperfect. Hence, the simulation itself is not required to be perfect, but simply an accurate representation of the electron-specimen interaction. The best simulation is arbitrary to some extent, and one could say that the best possible simulation would require an infinite number of calculations. By taking advantage of image inpainting and efficient parameter selection, it is possible to significantly reduce the number of required calculations and allow low-end machines to run higher resolution simulations with respect to the standard methods. Also, if it is intended that the user takes advantage of real CS-STEM, then the error in their simulation would unlikely be more than that of their real experimental data, given equivalent sampling regimes and conditions. This is mainly because noise and human error are redundant in a simulation. In addition to this, for those wishing to run real-time simulations alongside their experimental work, with further advancements in calculation techniques (such as the PRISM method in combination with a compressed model) it could soon be a possibility. Future works intend to take advantage of simulations in more ways than traditionally used as discussed in the introduction. By combining experiment with theory, it is expected that the efficiency of real CS-STEM will improve, and furthermore increase the quality of output through dictionary optimisation techniques. Faster simulations could offer a solution for real time compressive sensing STEM, and the potential to find optimal sampling strategies by real time microscope parameter simulation. SIM-STEM Lab version two is currently being worked on to try and further improve performance, and the solutions extended beyond annular dark field imaging.

# SIM-STEM Lab: Incorporating Compressed Sensing Theory for Fast STEM Simulation

Alex Robinson[1], Daniel Nicholls[1], Jack Wells[2], Amirafshar Moshtaghpour[1,3], Angus Kirkland[3,4], Nigel D. Browning[1,5,6]

[1] Department of Mechanical, Materials and Aerospace Engineering, University of Liverpool, Liverpool, L69 3GH, United Kingdom

[2] Distributed Algorithms Centre for Doctoral Training, University of Liverpool, Liverpool, L69 3GH, United Kingdom

[3] Correlated Imaging group, Rosalind Franklin Institute, Harwell Science and Innovation Campus, Didcot, OX11 0QS, United Kingdom.

[4] Department of Materials, University of Oxford, Oxford, OX2 6NN, United Kingdom

[5] Physical and Computational Science Directorate, Pacific Northwest National Laboratory, Richland, WA 99352, USA

[6] Sivananthan Laboratories, 590 Territorial Drive, Bolingbrook, IL 60440. USA

## Supplementary Material

### Supplementary Figures

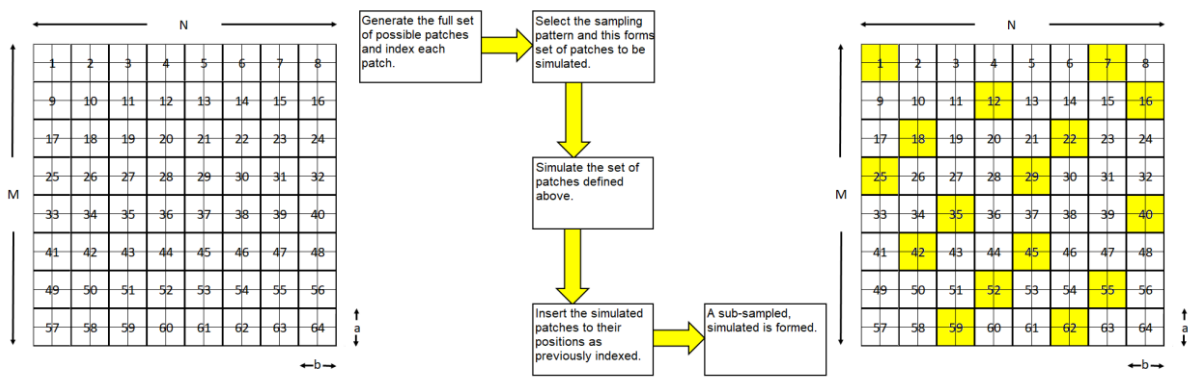

**Figure S1**. An image of size $[M \times N]$ pixels can be broken down into a set of patches of size $[a \times b]$ (Left). Each patch is then indexed and creates a vector of patches which can be individually simulated depending on the desired scan pattern and sampling percentage. The simulated patches are then restored back into their respective index position (right), and those which are not simulated are set to a value of zero. The workflow to form a subsampled, simulated image (middle).

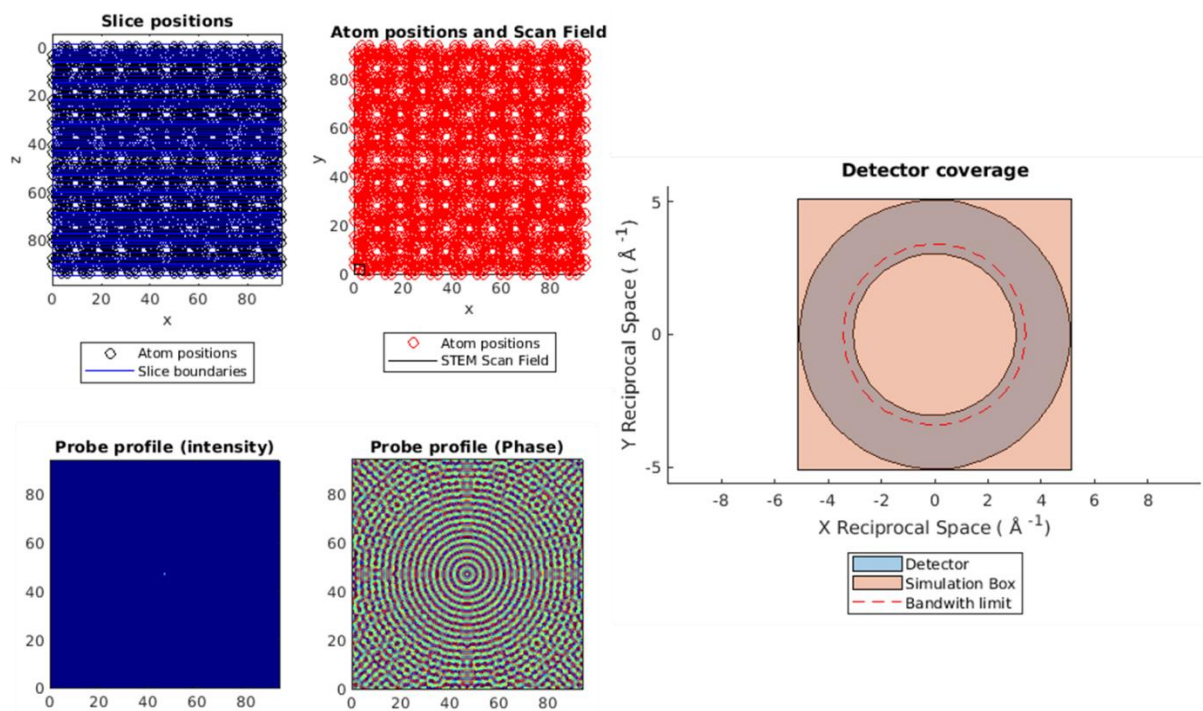

**Figure S2**. Details of the simulation probe, slicing and detector coverage parameters. The figures are provided through MULTEM in MATLAB. The rightmost figure shows the maximum reciprocal space vector limited to the edge of the detector. The left-hand side figures show details of the simulated STEM probe, atom positions, slices and scan field.

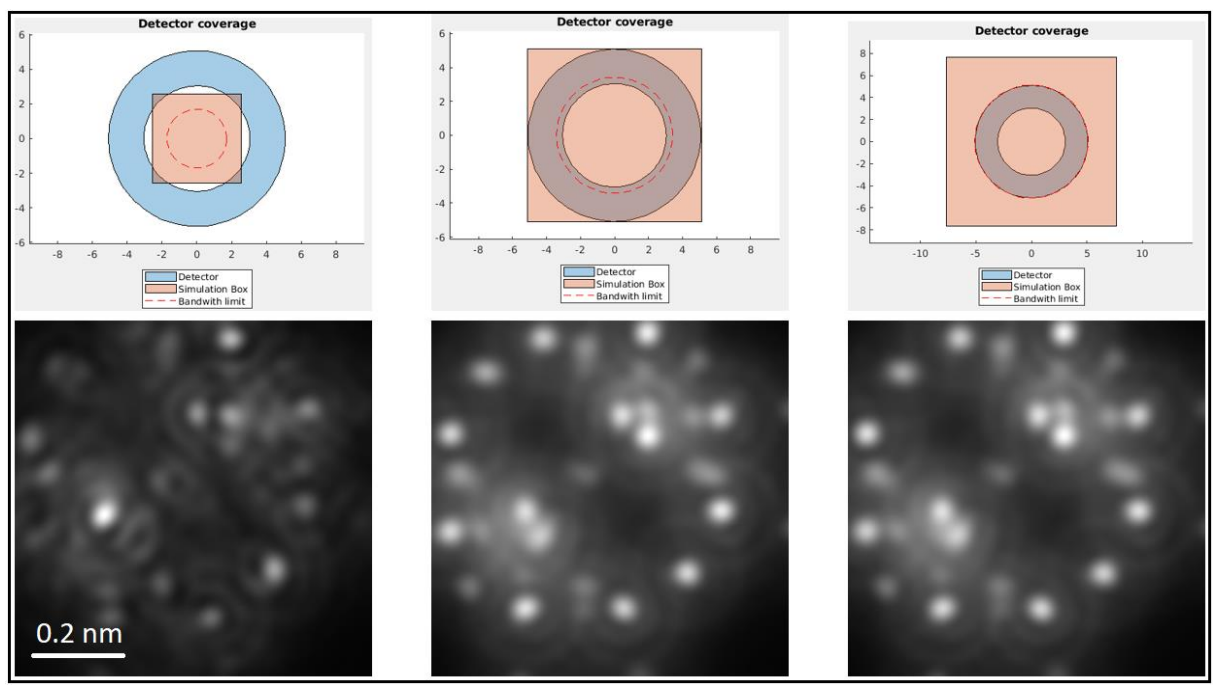

**Figure S3**. A demonstration of how the simulation box, or maximum reciprocal space vector changes the outcome of the simulation. When it is too small, the electrons which would have scattered to angles incident upon the detector would not contribute, and beyond the limit of the detector, the improvement in simulation is diminishing.

| Sampling Percentage (%) | Percentage Time to Completion With Respect to Reference Image (%) | μ SSIM (%) | σ SSIM (%) | μ PSNR (dB) | σ PSNR (dB) |
|---|---|---|---|---|---|
| 5 | 0.33 | 59.50 | 3.64 | 18.98 | 1.00 |
| 10 | 0.60 | 83.73 | 2.99 | 24.60 | 1.02 |
| 15 | 0.86 | 90.55 | 3.30 | 26.97 | 1.88 |
| 20 | 1.14 | 95.16 | 0.66 | 30.88 | 0.70 |
| 25 | 1.42 | 95.85 | 0.58 | 31.40 | 1.87 |
| 30 | 1.67 | 96.10 | 0.65 | 31.91 | 0.97 |
| 35 | 1.96 | 96.59 | 0.52 | 32.18 | 1.09 |
| 40 | 2.21 | 96.87 | 0.26 | 32.81 | 0.69 |

| Number of Frozen Phonon Configurations (arb) | Percentage Time to Completion With Respect to Reference Image (%) | μ SSIM (%) | σ SSIM (%) | μ PSNR (dB) | σ PSNR (dB) |
|---|---|---|---|---|---|
| 1 | 0.74 | 95.04 | 0.71 | 26.83 | 1.80 |
| 2 | 1.48 | 97.53 | 0.19 | 30.98 | 0.53 |
| 4 | 2.95 | 98.51 | 0.29 | 33.06 | 2.00 |
| 8 | 5.90 | 99.11 | 0.21 | 35.38 | 2.54 |
| 12 | 8.84 | 99.31 | 0.12 | 36.31 | 2.03 |
| 16 | 11.78 | 99.39 | 0.10 | 36.31 | 1.92 |
| 24 | 17.67 | 99.47 | 0.07 | 36.48 | 1.69 |
| 32 | 23.57 | 99.49 | 0.07 | 36.38 | 1.64 |

| Maximum Reciprocal Space Vector (Å$^{-1}$) | Percentage Time to Completion With Respect to Reference Image (%) | μ SSIM (%) | σ SSIM (%) | μ PSNR (dB) | σ PSNR (dB) |
|---|---|---|---|---|---|
| 1.37 | 0.88 | 0.69 | 0.00 | 10.04 | 0.00 |
| 2.73 | 1.14 | 90.81 | 0.82 | 25.11 | 0.92 |
| 4.10 | 2.40 | 98.14 | 0.33 | 32.12 | 2.01 |
| 5.46 | 3.69 | 98.77 | 0.25 | 34.13 | 2.12 |
| 6.83 | 9.26 | 98.91 | 0.17 | 35.04 | 1.68 |
| 8.19 | 13.60 | 98.93 | 0.17 | 35.18 | 1.69 |
| 9.56 | 16.84 | 98.93 | 0.18 | 35.19 | 1.84 |
| 10.92 | 15.63 | 98.93 | 0.18 | 35.20 | 1.86 |

**Figure S4**. The data used in figures 7-9.

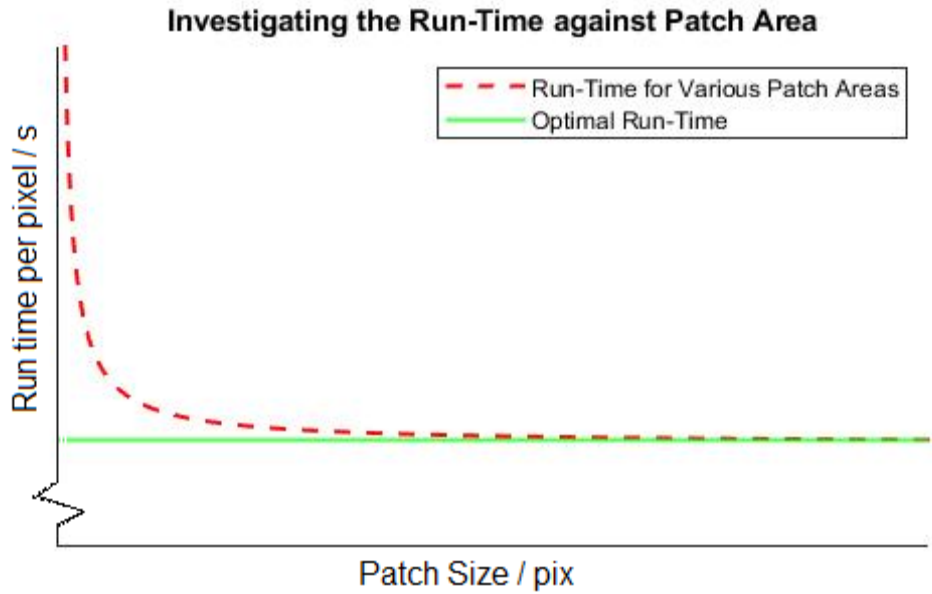

**Figure S5**. Empirical data based on the run-time of increasing patch areas used in the subsampling of simulations. Smaller patches are more inefficient due to the time it takes to transfer data from GPU to CPU memory. Ideally, we would expect a linear relation between the sampling percentage and the time it takes to run the subsampled simulation.